\documentstyle{article}
\newcommand{\be}{\begin{equation}}
\newcommand{\ee}{\end{equation}}
\newcommand{\bea}{\begin{eqnarray}}
\newcommand{\eea}{\end{eqnarray}}
\newcommand{\ba}{\begin{array}}
\newcommand{\ea}{\end{array}}

\def\bbox{{\,\lower0.9pt\vbox{\hrule \hbox{\vrule height 0.2 cm
\hskip 0.2 cm \vrule height 0.2 cm}\hrule}\,}}
\newcommand{\dsl}{\pa \kern-0.5em /}

\newcommand{\nn}{\nonumber \\}

\newcommand{\EQ}{\begin{equation}}
\newcommand{\EN}{\end{equation}}

\def\bbox{{\,\lower0.9pt\vbox{\hrule \hbox{\vrule height 0.2 cm
\hskip 0.2 cm \vrule height 0.2 cm}\hrule}\,}}

\newcommand{\pa}{\partial}

\def\bfO{\mbox{\boldmath $\Omega$}}

\def\today{\ifcase\month\or
  January\or February\or March\or April\or May\or June\or
  July\or August\or September\or October\or November\or December\fi
 \space\number\day, \number\year}

\input amssym.def
\input amssym.tex
\font\mybb=msbm10 at 10pt
\def\bb#1{\hbox{\mybb#1}}

\def\bE {\bb{E}}

\def\bfO{\mbox{\boldmath $\Omega$}}

\begin{document}


\begin{titlepage}
\vfill
\begin{flushright}
DAMTP-2001-67\\
hep-th/0107228\\
\end{flushright}

\vskip 1cm
\begin{center}
\baselineskip=16pt
{\Large\bf THE FIRST LAW OF BLACK BRANE MECHANICS}
\vskip 0.3cm
{\large {\sl }}
\vskip 10.mm
{\bf  ~Paul K. Townsend and Marija Zamaklar} \\
\vskip 1cm
{\small
DAMTP, University of Cambridge, \\
Centre for Mathematical Sciences,
Wilberforce Road, \\
Cambridge CB3 0WA, UK\\
}
\end{center}
\vskip 1cm

\begin{center}
{\bf ABSTRACT}
\end{center}
\begin{quote}

We obtain ADM and Komar surface integrals for the energy density, tension
and angular momentum density of stationary $p$-brane solutions of
Einstein's equations. We use them to derive a Smarr-type formula for the
energy density and thence a first law of black brane mechanics. The
intensive variable conjugate to the worldspace p-volume is an `effective'
tension that equals the ADM tension for uncharged branes, but vanishes
for isotropic  boost-invariant charged branes.
   
\end{quote}
\end{titlepage}

\setcounter{equation}{0}
\section{Introduction}

The status of energy in General Relativity is rather special because, for a
general spacetime, there is no coordinate-independent meaning
to the energy in a given local region of a spacelike hypersurface. However,
if the spacetime is asymptotically flat then it is possible to define a {\it
total} energy, which can be expressed as an integral at spatial infinity:
the ADM integral \cite{ADM}. Consider an asymptotically
flat spacetime, of $d=1+n$ dimensions, for which the asymptotic form of the
metric, in {\it cartesian} coordinates $x^M=(t,x^i)$ ($i=1,\dots,n$) is
\be
g_{MN}\sim \eta_{MN} + h_{MN}
\ee
where $\eta$ is the Minkowski metric and $h$ is a perturbation with the
appropriate fall-off at spatial infinity needed to make the energy
integral converge. We shall suppose that the action takes the
form\footnote{This corresponds to a choice of units for which $8\pi G=1$,
where $G$ is the $d$-dimensional Newton constant.}
\be\label{action1}
I = {1\over 2}\int d^dx\, \sqrt{-\det g}\,  R \ + I^{(mat)}\, .
\ee
The matter stress-tensor is
\be
T^{(mat)}_{MN} = - {2\over \sqrt{-\det g}}\ {\delta I^{(mat)}\over \delta
g^{MN}}\, ,
\ee
and the Einstein field equation is
\be
G_{MN}= T_{MN}^{(mat)}\, .
\ee
The ADM energy integral in these conventions is
\be\label{ADM}
E = {1\over2}\oint_\infty dS_i
\left(\partial_j h_{ij} - \partial_i h_{jj}\right)\, .
\ee

In the special case of a stationary asymptotically flat spacetime, for which
there exists a normalized timelike Killing vector field $k$, the energy in
any volume $V$ with boundary $\partial V$ can be expressed as the Komar
integral \cite{Komar}
\be\label{komar}
E(V)= -\left({n-1\over 4(n-2)}\right) \int_{\partial V} dS_{MN} D^M
k^N\, .
\ee
where $dS_{MN}$ is the co-dimension $2$ surface element on the
$(n-1)$-dimensional boundary $\partial V$ of $V$. By taking $V$ to be a
spacelike Cauchy surface $\Sigma$ with boundary at spatial infinity, we get
an alternative expression for the total energy $E$ in which $k$ need be
only {\sl asymptotically} Killing. Taking $\Sigma$ to be a surface of
constant $t$, in coordinates for which $k=\partial/\partial t$, we thus find
that
\be\label{goo}
E= -\left({n-1\over 2(n-2)}\right) \oint_\infty dS_i\, \partial_i\, h_{00}\,
,
\ee
which shows that the energy can be read off from the asymptotic behaviour of
the $g_{00}$ component of the metric.

One aim of this paper is to generalize these surface integrals to energy
densities of `p-brane spacetimes', which are not asymptotically flat but
rather `transverse asymptotically flat', and to demonstrate their
equivalence. One difference between the $p=0$ and $p>0$ cases is that
for $p>0$ the energy is replaced by a stress-tensor density, which has
energy per unit p-volume ${\cal E}$ (the energy density) and $p$ tensions as
its $(p+1)$ diagonal entries. For an isotropic brane all $p$ tensions are
equal, and equal by definition to the brane tension ${\cal T}$.  Although
isotropy it is not really  necessary to our analysis we will mostly consider
isotropic branes for the sake of simplicity. It should be noted that the
energy density is sometimes referred to in the brane literature as the
`tension', but it actually equals the tension only for boost-invariant
isotropic branes, examples of which are provided by many extremal brane
spacetimes. Here we wish to consider the more general case of
non-extremal, and non boost-invariant, brane spacetimes, so we must
distinguish between tension ${\cal T}$ and energy density ${\cal E}$.

The ADM integral for the energy of transverse asymptotically flat
spacetimes was first obtained by Deser and Soldate \cite{deser} for $p=1$,
using the method introduced by Abbott and Deser \cite{AD}, and the result
for
general $p$ was given in \cite{Stelle}. We will present here the extension
of
the argument of \cite{deser} to general $p$. We also obtain ADM-type
integrals
for the tension and angular momentum. We then use these results to find the
corresponding Komar-type integrals for these quantities, generalizing the
standard covariant Komar surface integrals of the $p=0$ case. For $p=0$
these
integrals are unique up to normalization, but this feature does not extend
to
$p>0$, which is why we need to start from the ADM integrals.  A corollary of
our results is that when $p>0$ {\it neither the energy density nor the brane
tension can be read off from the $g_{00}$ component of the metric}, an
observation that seems to have first been made (for the energy density) by
Lu
\cite{Lu} in the context of a special class of brane spacetimes.

Another aim of this paper is to derive a first law for black branes
analogous to the first law of black hole mechanics \cite{BCH}. For black
holes the first law can be derived from a generalization of the Smarr
formula
for the energy of a black hole spacetime \cite{Smarr}, which can itself be
deduced from the Komar integrals for black hole energy and angular momentum
\cite{BCH}. The covariant surface integrals for the energy and angular
momentum densities of brane spacetimes therefore constitute a natural
starting point for a derivation of the first law of black brane mechanics.
There is one version of this law in which the intensive variables conjugate
to the surface horizon $\kappa$, the angular velocities of the horizon
$\bfO_H$ and the (electric) potential of the the horizon $\Phi_H$ are
{\it densities}. This law relates a change of the energy density
to changes in other densities such as the angular momentum densities ${\bf
{\cal J}}$ and the (electric) charge density ${\cal Q}$, and it turns out
to take the form
\be
d{\cal E} = \kappa d{\cal A}_{eff} + \bfO_H\cdot d{\bf {\cal J}} + \Phi_H
d{\cal Q} \, ,
\ee
where ${\cal A}_{eff}$ is an `effective horizon area'. Because of some
simplifying assumptions that we make, there are brane spacetimes to which
our
derivation of this law does not apply (athough the law itself may remain
valid). These would include, for example, 5-branes of D=11 supergravity
carrying both 5-brane and 2-brane charge. More significantly, we consider
only translationally-invariant brane spacetimes, although our covariant surface
integrals are valid for brane spacetimes that are only {\it asymptotically}
translationally invariant. Even so, our analysis applies to many cases of
interest, including that of the rotating non-extremal membrane solutions of
D=11 supergravity obtained in \cite{cvetic}. Recent work of 
Traschen and Fox \cite{TF} in which a version of the first law of
black brane mechanics is derived by Hamiltonian methods appears to go 
some way towards relaxing these restrictions\footnote{We became aware
of this paper after an earlier submission to the archives of the work
reported here; the precise connection to the work reported here is not
yet clear to us.}.

There is another version of the first law that applies
to toroidally-compactified branes in which a change in the {\it total}
energy
$E$, integrated over the p-dimensional worldspace, is related to changes in
the total horizon area $A$, total angular momenta $J$ and total charge $Q$,
as
for black holes. This is hardly surprising because a toroidally-compactified
black brane can be viewed as a black hole in the lower dimension. From the
higher-dimensional point of view, however, the worldspace p-volume $V_p$ is
a
new extensive variable. If we call its conjugate variable the `effective'
tension ${\cal T}_{eff}$ then the first law becomes
\be\label{intversion}
dE = \kappa dA + \bfO_H \cdot {\bf J} + \Phi_H dQ + {\cal T}_{eff} dV_p\, .
\ee
One might expect that the effective tension appearing in this law should
equal the ADM tension ${\cal T}$, but this turns out to be the case only for
uncharged branes. For charged branes we find that
\be
{\cal T}_{eff} = {\cal T} - \Phi_H {\cal Q}\, ,
\ee
so that the effective tension is reduced by a pressure due to the gauge
fields produced by the charge on the black brane.Of course, the variable
$V_p$ also has a `lower dimensional' interpretation as a scalar field
expectation value, in which case its conjugate variable ${\cal T}_{eff}$
could be called a `scalar charge' \cite{gwg}. An extension of the first law
to accomodate scalar fields and their scalar charges has been presented
previously by Gibbons et al. \cite{GKK}; the novelty of our formulation of
this version of the first law is the interpretation of scalar charge in
terms of ADM tension. 

For isotropic brane solutions of supergravity theories there is a
bound on the energy density \cite{GHT}, which is saturated by
supersymmetric solutions. We shall show that this energy bound implies
the bound ${\cal T} \le {\cal E}$, which is also saturated by
supersymmetric solutions. Thus, supersymmetric isotropic branes are
boost-invariant. The energy bound also implies the bound
${\cal T}_{eff} \ge 0$, which is again saturated by supersymmetric
solutions. This can be
understood from a worldvolume perspective; the effective tension is
the vacuum energy, which has ${\cal T}$ as a gravitational
contribution and the term proportional to ${\cal Q}$ as an
`electrostatic' contribution, the two cancelling in the supersymmetric
case.

\section{ADM integrals}

Our first task is to obtain the ADM integrals for brane energy density,
tension and angular momentum. We shall do this using a generalization of the
method of Abbott and Deser \cite{AD}. This method can be used to define the
energy as a surface integral at infinity in any spacetime that is asymptotic
to some background spacetime admitting a timelike Killing vector field. Here
we shall apply it to spacetimes that are transverse asymptotically flat, so
that the background spacetime is a Minkowski spacetime of $D= (n+p+1)$
dimensions, to
which the full metric $g_{MN}$ is asymptotic at transverse spatial infinity.
We may write
\be
g_{MN}= \eta_{MN} + h_{MN}
\ee
where $h_{MN}$ is a (not necessarily small) perturbation about the Minkowski
background metric $\eta$. We can now rewrite the Einstein equation as
\be\label{lin}
G^{(lin)}_{MN} = T^{(mat)}_{MN} + \tau_{MN}
\ee
where $G^{(lin)}_{MN}$ is the linearized Einstein tensor and $\tau^{MN}$ is
the gravitational stress pseudo-tensor; i.e., the stress tensor of the
gravitational field relative to the $D$-dimensional Minkowski background.
Let
\be
H_{MN} = h_{MN} - {1\over2}\eta_{MN}h\, ,
\ee
where $h= h^M{}_M$, and define
\be
K^{MPNQ} = {1\over2}\left[\eta^{MQ}H^{NP} +
\eta^{NP}H^{MQ} -\eta^{MN}H^{PQ} -
\eta^{PQ}H^{MN}\right]\, .
\ee
Note that $K$ has the algebraic symmetries of the Riemann tensor.
We can now rewrite the linearized Einstein tensor, in {\it cartesian}
coordinates, as
\be\label{Glin}
G_{(lin)}^{MN} = \partial_P\partial_Q K^{MPNQ}\, .
\ee
Using this, we can  rewrite the Einstein equation (\ref{lin}) in
the form
\be\label{Einequiv}
\partial_P\partial_Q K^{MPNQ} = T^{MN}\, ,
\ee
where
\be
T^{MN} = T_{(mat)}^{MN} +\tau^{MN}
\ee
is the `total' stress pseudo-tensor. Consistency of (\ref{Einequiv})
requires that 
\be\label{conservation}
\partial_M T^{MN} =0\, .
\ee

Let $X^M=(t,y^m,x^i)$ ($i=1,\dots, n$; $m=1,\dots,p$) be cartesian
coordinates for the background Minkowski spacetime, so that
\be
ds^2(\eta) = -dt^2 + \sum_{m=1}^p dy^m dy^m + \sum_{i=1}^n dx^i dx^i\, .
\ee
Note that
\be
\bar k= {\partial\over \partial t}\, , \qquad \bar \ell_{(m)}={\partial
\over
\partial y^m}\, ,
\ee
are commuting Killing vector fields of this background metric, which we may
write collectively as $\bar k_{(\mu)}$, $\mu=0,1\dots p$. We will assume
that
the full metric $g$ is asymptotic to $\eta$ as $|x|\rightarrow \infty$; it
is
thus `transverse asymptotically' flat, stationary and translationally
invariant (in brane directions). Equivalently, we assume that $T^{MN}$
vanishes in the limit $|x|\rightarrow 0$ (sufficiently rapidly for
convergence of the energy integral to be discussed below). Because of
the condition (\ref{conservation}), the currents
\be
J^M_{(\mu)} = T^M{}_N \bar k_{(\mu)}^N
\ee
are divergence-free (where indices are raised or lowered with the background
Minkowski metric).
We may also construct an analogous set of p-form currents:
\bea
A_{(0)}^{P_1\dots P_p} &=& \bar\ell_{(1)}^{[P_1} \dots
\bar\ell_{(p)}^{P_p]}\nn
A_{(m)}^{P_1\dots P_p} &=& \bar \ell_{(1)}^{[P_1} \dots
\bar k^{P_m} \dots
\bar \ell_{(p)}^{P_p]}\, ,
\eea
where $\bar k$ replaces $\bar \ell_{(m)}$ in the second expression. Because
the Killing vector fields commute, the p-form currents $A_{(\mu)}$ are also
divergence free, in the sense that (in cartesian coordinates)
\be
\partial_S A_{(\mu)}^{SR_1\dots R_{p-1}} =0\, \qquad (\mu=0,1,\dots,p).
\ee

We now define the brane stress tensor density as
\be
\theta_{\mu\nu} = -\int_\Sigma dS_{MP_1\dots P_p} \,
J_{(\mu}^{M} A_{\nu)}^{P_1\dots P_p}
\ee
where $dS_{MP_1\dots P_p}$ is the co-dimension $(2+p)$ surface element for
$\Sigma$, normal to $\bar k$ and the $p$ vector fields $\{\bar\ell\}$. In
standard cartesian coordinates for the background Minkowski metric, we have
\be
\theta_{\mu\nu} = \int\! d^n x\, T_{\mu\nu}
\ee
where $T_{\mu\nu}$ are the worldvolume components of $T_{MN}$, and the
integral is over the transverse Euclidean n-space. Clearly,
$\theta_{\mu\nu}$
can be written as a surface integral at transverse spatial infinity only if
it is independent of the worldvolume coordinates $y^\mu= (t,y^m)$, since any
surface integral will have this property as a consequence of the assumption
of asymptotic time and translational invariance. But this condition is met
only if $J_{(\mu}\wedge A_{\nu)}$ is a divergence-free $(p+1)$-current, and
this requires both
\be
\left({\cal L}_{\bar k} T\right)_{MN} =0
\ee
and
\be\label{traninv}
\left({\cal L}_{\bar \ell_{(m)}}T\right)_{MN} =0 \qquad (m=1,\dots, p)\, .
\ee
These conditions are quite severe as they effectively restrict us to
spacetimes that are {\it everywhere} both stationary and
translationally-invariant, rather than just asymptotically so. If we
consider
only the component $\theta_{00}$ then we need only require
(\ref{traninv}) but this still restricts us to translationally-invariant
brane spacetimes. To circumvent this restriction we shall impose periodicity
in the brane directions, with total p-volume $V_p$, and define
\be
\langle A \rangle = {1\over V_p}\int d^py \, A
\ee
for any function $A$ on the background Minkowski spacetime. The average
energy density is then
\be
{\cal E} \equiv \langle \theta_{00} \rangle = \int d^n x\, \langle T^{00}
\rangle\, .
\ee
For stationary spacetimes we may similarly define the
average\footnote{In the non-isotropic case
this is also an average over brane directions.} tension ${\cal T}$ as
\be\label{ET}
{\cal T} \equiv -{1\over p}\sum_m \langle \theta_{mm} \rangle =
-{1\over p}\sum_m \int d^n x\, \langle T^{mm} \rangle\, .
\ee
An analogous procedure yields the expression
\be
{\cal L}^{ij} = 2\int d^nx\, x^{[i}\langle T^{j]0}\rangle\, .
\ee
for the average p-brane angular momentum 2-form density.

We now aim to rewrite these transverse-space integrals as surface
integrals using (\ref{Einequiv}). Having done so we may then omit the
averaging because of the translational invariance at transverse spatial
infinity. For the energy density we thus find that
\be\label{Kten}
{\cal E} = \oint_\infty dS_i \, \partial_j  K^{i0j0}\, ,
\ee
and for the angular momentum 2-form density we find that
\be\label{Kang}
{\cal L}^{ij} = 2 \oint_\infty dS_\ell \left[ K^{ij0\ell} +
x^{[i}\partial_k  K^{j]\ell 0k} + x^{[i}\partial_0
K^{j]00\ell}\right]\, ,
\ee
where the integrals are over an $(n-1)$-sphere at transverse spatial
infinity. These expressions are the basis for the ADM-type formulae for
brane energy and angular momentum.
Specifically, the surface integral (\ref{Kten}) is equivalent to
\be\label{pADM}
{\cal E} = {1\over2}\oint_\infty dS_i \left(\partial_j  h_{ij} - \partial_i
 h_{jj} -\partial_i h_{mm}\right)\, .
\ee
This is the result given in \cite{deser,Stelle}. For asymptotically
stationary spacetimes, for which time derivatives vanish at infinity, the
integral (\ref{Kang}) is similarly equivalent to\footnote{See
\cite{Ash1} for a discussion of angular momentum in General Relativity.}
\be\label{Kang2}
{\cal L}^{ij} = \oint_\infty dS_\ell \left[ 2\delta^{\ell [i} h^{j]0}
-  x^{[i}\partial_\ell h^{j]0} +  x^{[i}\partial^{j]} h^{\ell 0}
\right]\, .
\ee
This result implies that the angular momentum density can be
read off from the coefficient of $h_{0i}$ according to the formula
\be
h_{0i} = {{\cal L}_{ij}x^j\over vol_{(n-1)} r^n}
\ee
where $vol_k$ is the volume of the unit $k$-sphere.

For future use we note that for constants $\Omega_{ij}=-\Omega_{ji}$ we have
\be
{1\over2}\Omega_{ij}{\cal L}^{ij} = \oint_\infty dS_i \left[ m_{i,j}
h_{0j} - {1\over2}\left(h_{0i,j} - h_{0j,i}\right)m_j\right]
\ee
where
\be\label{asymm}
m_i = \Omega_{ij}x^j\, ,
\ee
which are the components of a rotational Killing vector field $m$ of
$\bE^n$.  

\section{Komar integrals}

The ADM-type surface integral (\ref{pADM}) can be written as the following
bulk integral 
\be\label{ADMen}
{\cal E} = {1\over2}\int d^nx\,
\left[h_{ij,ij} - h_{jj,ii} -  h_{mm,ii}\right]\, ,
\ee
where repeated indices are summed.
This assumes that the metric perturbation is not so large in the interior
that it changes the topology of space; otherwise we cannot view the metric
perturbation as a field on Minkowski spacetime. However, as long as we
express any final result as a surface integral at infinity,
this restriction can make no
difference to the result. We may now use the Einstein equations in the form
(\ref{lin}) to deduce alternative bulk energy integrals, some of which may
then be re-expressed as surface integrals; this is how one gets from the ADM
energy integral (\ref{ADM}) to the Komar formula (\ref{goo}). We shall now
explain this point in some detail for the general $p\geq 0$ case.

The equations (\ref{lin}) are equivalent to the Pauli-Fierz  equations
\be
\Box h_{MN} + h_{,MN} -2h_{(M,N)} = -2\left(T_{MN} - {1\over
D-2}\eta_{MN}T\right)
\ee
where $T=\eta^{MN}T_{MN}$. For a {\it time-independent metric perturbation}
these equations imply that
\be\label{fromPF}
(D-3) \left[\langle h_{ij}\rangle_{,ij} - \langle h_{jj}\rangle_{,ii}
-\langle h_{mm}\rangle_{,ii}\right] = -
\left[(D-2)
\nabla^2 \langle h_{00}\rangle + 2\left(\langle T_{ii}\rangle  +\langle
T_{mm}\rangle\right)\right]\, .
\ee
Integrating both sides over the n-dimensional transverse Euclidean
space we deduce (as long as $D>3$) that
\be\label{tensionp}
{\cal E} =  -\left({D-2\over 2(D-3)}\right) \oint_\infty dS_i\, \partial_i
h_{00} + {p\over D-3} {\cal T} - {1\over D-3}\int  d^nx\, \langle T_{ii}
\rangle\, ,
\ee
where ${\cal E}$ is the ADM energy density (\ref{ADMen}) and
${\cal T}$ is the tension, as given in (\ref{ET}). Now, the
conservation condition (\ref{conservation}) implies that
\be\label{cons}
\partial_i \langle T_{ij} \rangle =0\, ,
\ee
and hence that
\be\label{lemma}
\int d^n x\, \langle T_{ii} \rangle = \oint_\infty dS_i\, (x^j T_{ij})
=0\, ,
\ee
where the last equality follows from the assumption that $T_{ij}$ vanishes
at
transverse spatial infinity. Thus, the $T_{ii}$ term in (\ref{tensionp})
does
not contribute. Consequently, (\ref{tensionp}) reduces to (\ref{goo}) for
$p=0$. For $p>0$, however, {\it the tension contributes to the energy
density}
and we need to rewrite it too as a surface term. Using the Pauli-Fierz
equation again we deduce, after further use of (\ref{lemma}), that
\be\label{altten}
{\cal T} = -\left({1 \over 2p(n-2)}\right)\oint dS_i\,
\partial_i [p\, h_{00} - (n+p-2)h_{mm}]\, ,
\ee
where we have again omitted the averaging since this makes no difference
at transverse spatial infinity.  Combining this
formula with (\ref{tensionp}), we obtain the alternative energy surface
integral (summed over the index $m$)
\be\label{pgoo}
{\cal E} = -\left({1\over 2(n-2)}\right)
\oint_\infty dS_i\, \partial_i \left[ (n-1)  h_{00} -  h_{mm}\right]
\, .
\ee
This is equivalent to the formula given in \cite{Lu}, but now derived for a
much more general class of spacetimes (even within the subclass of
translationally invariant ones). In addition (\ref{altten}) provides a
similar surface integral for the tension ${\cal T}$. Note that
for isotropic boost invariant p-brane spacetimes we have $h_{mm} = -h_{00}$
for each $m$, so that ${\cal T} ={\cal E}$, as claimed earlier.

We now aim to rewrite these surface integrals in covariant form.
Let $k$ and $\ell_{(m)}$ be vector fields that are asymptotic to
$\bar k$ and $\bar\ell_{(m)}$ as $|x|\rightarrow \infty$. Then,
as may be verified,
\bea\label{Tcov}
{\cal E} &=& -\left({1\over 2(n-2)}\right)
\oint_\infty dS_{MN P_1\dots P_p}\bigg[(n-1) (D^M k^N)
\ell_{(1)}^{P_1}\cdots
\ell_{(p)}^{P_p} \nn
&&\ \qquad \qquad \qquad \qquad \qquad \qquad
+\ k^N D^M (\ell_{(1)}^{P_1}\cdots
\ell_{(p)}^{P_p})\bigg]\, ,
\eea
where $D$ is the usual gravitational covariant derivative associated
with the full metric $g$. Similarly, (for $p>0$) we have
\bea\label{Tcov2}
{\cal T} &=& -\left({1\over 2p(n-2)}\right)
\oint_\infty dS_{MN P_1\dots P_p}\bigg[ p\, (D^M k^N)
\ell_{(1)}^{P_1}\cdots
\ell_{(p)}^{P_p} \nn
&&\ \qquad \qquad \qquad \qquad \qquad
+\ \left(n+p-2\right)\, k^N D^M \left(\ell_{(1)}^{P_1}\cdots
\ell_{(p)}^{P_p}\right)\bigg]\, .
\eea

The angular momentum per unit p-volume ${\cal J}$ associated with a
rotational Killing vector field $m$ may be similarly expressed as a
covariant surface integral. Let $m$ have the asymptotic form (\ref{asymm})
in
cartesian coordinates at transverse spatial infinity, and define
\be
{\cal J}(m) = {1\over 2}\Omega_{ij}L^{ij} \, ,\qquad \left(m\sim
\Omega^i{}_jx^j{\partial\over\partial x_i}\right).
\ee
It may be verified that
\be
{\cal J}(m) = {1\over2} \oint_\infty dS_{MNP_1\dots P_p}
(D^Mm^N)\ell_{(1)}^{P_1}
\cdots
\ell_{(p)}^{P_p}\, .
\ee

\section{Smarr-type formula}

We now aim to use the covariant surface integrals for brane tension and
angular momentum to derive an analog for branes of the Smarr formula for
the mass of a stationary black hole \cite{Smarr}. We will suppose that a
stationary p-brane spacetime is regular on and outside a Killing horizon of
the Killing vector field
\be\label{xi}
\xi = k + \bfO_H \cdot {\bf m}
\ee
where $k$ is the unique normalized Killing vector field that is timelike
at transverse spatial infinity (normalized such that $k^2 \rightarrow -1$)
and ${\bf m}=(m_1,\dots)$ is a set of {\it commuting} rotational
Killing vector fields, normalized such that their orbits have length $2\pi
R$ as $R\rightarrow \infty$, where $R$ is distance from the horizon; the
constants $\bfO_H$ are the corresponding components of the
angular velocity of the horizon. Now use (\ref{xi}) to rewrite the
expression
(\ref{Tcov}) as
\bea\label{Tpnew}
{\cal E} &=& -{1\over 2(n-2)}\ \oint_\infty dS_{MNP_1\dots
P_p}\Big\{ (n-1) (D^M\xi^N)
\ell_{(1)}^{P_1}\dots \ell_{(p)}^{P_p} \nn
&&\qquad +\ \xi^N D^M\left(\ell_{(1)}^{P_1}\dots
\ell_{(p)}^{P_p} \right)\Big\} +
\left({n-1\over n-2}\right) \bfO_H\cdot{\bf {\cal J}}\nn
&&\ +\ {1\over 2(n-2)} \bfO_H\cdot \oint_\infty dS_{MNP_1\dots P_p} {\bf
m}^N
D^M\left(\ell_{(1)}^{P_1}\dots \ell_{(p)}^{P_p}\right) \, ,
\eea
where 
\be
{\bf {\cal J}} = {1\over2} \oint_\infty dS_{MNP_1\dots P_p} (D^M{\bf
m}^N)\ell_{(1)}^{P_1}
\cdots \ell_{(p)}^{P_p}\, .
\ee
By means of Gauss's law, we can rewrite the surface integral at infinity for
${\cal E}$ as the sum of a surface integral over the horizon and a bulk
integral over the region bounded by the horizon and transverse spatial
infinity on an $n$-surface $\Sigma$ having $k$ as one of its
$(p+1)$ normals. In
doing so we shall assume that the vector fields $\ell_{m}$ are
{\it everywhere} Killing, and not just asymptotically Killing,
which means that {\it we now make the restriction to brane
spacetimes that are translationally invariant as well as
stationary}. This means that there are $p$ translational Killing vector
fields
$\{\ell\}$ in addition to the timelike Killing vector field $k$, and we
assume that all are mutually commuting.

We begin by spelling out the
conditions that the rotational Killing vector fields ${\bf m}$ must satisfy,
corresponding to the assumption that they generate rotations in the
{\it transverse} space. We assume that {\it each of the vector fields
$\{\ell\}$ commutes with ${\bf m}$}. We shall also assume that the
rotational Killing vector fields ${\bf m}$ satisfy
\be\label{restrictm}
{\bf m}^MdS_{MN_1\dots N_p}\equiv 0\, ,
\ee
where $dS_{MN_1\dots N_p}$ is the volume element of $\Sigma$. This condition
is automatically satisfied for $p=0$ as long as $\Sigma$ is spacelike; if,
for $p>0$, the Killing vector fields $\{\ell\}$ are normal to $\Sigma$ then
it is equivalent to ${\bf m}\cdot \ell_{(m)}=0$. Given these restrictions on
${\bf m}$, an application of Gauss's law yields
\be
\oint_\infty dS_{MNP_1\dots P_p}\,  {\bf m}^N
D^M\left(\ell_{(1)}^{P_1}\dots \ell_{(p)}^{P_p}\right) =
\oint_H dS_{MNP_1\dots P_p}\, {\bf m}^N
D^M\left(\ell_{(1)}^{P_1}\dots \ell_{(p)}^{P_p}\right)\, .
\ee
An application of Gauss's law to the other surface integral on the right
hand side of (\ref{Tpnew}), yields both an integral over the horizon
and a bulk integral over $\Sigma$. Rewriting the latter in terms of
the matter stress tensor by means of the Einstein equation
\be\label{EinR}
R_{MN} = T^{(mat)}_{MN} - {1\over D-2} \, g_{MN} T^{(mat)}
\qquad \left(T^{(mat)}\equiv g^{MN}T^{(mat)}_{MN}\right)\, ,
\ee
we deduce that
\bea\label{newtension}
{\cal E} &=& - {1\over 2(n-2)}\, \oint_H dS_{MNP_1\dots P_p} \left\{ (n-1)
(D^M\xi^N) \ell_{(1)}^{P_1}\dots \ell_{(p)}^{P_p} +
\xi^ND^M\left(\ell_{(1)}^{P_1}\dots
\ell_{(p)}^{P_p}\right)\right\}\nn
&&+\  \left({n-1\over n-2}\right) \bfO_H \cdot  {\bf {\cal J}}\
+\ {1\over 2(n-2)} \bfO_H\cdot \oint_H dS_{MNP_1\dots P_p}\, {\bf m}^N
D^M\left(\ell_{(1)}^{P_1}\dots \ell_{(p)}^{P_p}\right)\nn
&& +  \ {\cal E}^{(mat)}\, ,
\eea
where
\bea\label{emat}
{\cal E}^{(mat)} &=& {1\over (n-2)} \int_\Sigma dS_{MNP_1\dots
P_{p-1}}\,\bigg\{ \ell_{(1)}^{P_1}\dots
\ell_{(p-1)}^{P_{p-1}}\ell_{(p)}^N\bigg[
(n-1)\left(T^{(mat)}\right)^M{}_Q\,
\xi^Q \nn
&&-\ \left(T^{(mat)}\right)\xi^M\bigg] -\ p\ \ell_{[(1)}^{P_1}\dots
\ell_{(p-1)}^{P_{p-1}}\ell_{(p)]}^Q\
\left(T^{(mat)}\right)^M{}_Q\, \xi^N\bigg\}\, .
\eea
For the moment we will assume that the matter stress tensor vanishes.
This means, in particular, that we are now
considering uncharged black branes since a charge would produce
`electric' or `magnetic' fields with a non-vanishing stress
tensor; we will consider the case of electrically-charged branes in the
following section. 

Given that ${\cal E}^{(mat)}$ vanishes, we have only to
simplify and interpret the horizon integrals in (\ref{newtension}) to obtain
a brane analogue of the Smarr formula for neutral black branes. On the
horizon, the orientation of the surface element $dS_{MNP_1\dots P_p}$ is
determined by $\xi$ itself,
$p$ other normals $\zeta_{(m)}$  and one other {\it null} vector $n$ which
we may choose such that
\be
\xi\cdot n =-1\, .
\ee
Since $\xi$ is tangent to the horizon, as well as normal to it, it must be
orthogonal to the other $p$ normals, so
\be
\xi\cdot \zeta_{(m)} =0\, .
\ee
Given that the horizon surface area element (per unit p-volume) has
magnitude $d{\cal A}$, we then have
\be\label{horsurfel}
dS_{MNP_1\dots P_p} = \left[(p+2)!\right] v^{-1}(\zeta)\, d{\cal A}\
\xi_{[M} n_N \zeta^{(1)}_{P_1}\dots \zeta^{(p)}_{P_p]}\, ,
\ee
where  
\be
v(\zeta) = |\zeta_{(1)}\wedge \dots \wedge \zeta_{(p)} |\, .
\ee
{}For a generic p-brane spacetime we may choose
\be\label{choice}
\zeta_{(m)} =\ell_{(m)}\, ,
\ee
although this will not be possible in those cases in which one or more of
the
vector fields $\ell_{(m)}$ has a fixed point on the horizon, as happens for
certain extremal $p$-brane solutions. However, such non-generic $p$-brane
spacetimes can be considered as limits of generic spacetimes, so we shall
assume here that the choice (\ref{choice}) is possible. In this case
$v(\zeta)$ equals $v(\ell)$, which is the ratio of the worldspace volume on
the horizon to the worldspace volume $V_p$ at infinity.

Given that
\be
(\xi\cdot D \xi)^M = \kappa \xi^M
\ee
on the horizon, for surface gravity $\kappa$, we then find that
\be
dS_{MNP_1\dots P_p}(D^M\xi^N)\ell_{(1)}^{P_1} \dots \ell_{(p)}^{P_p}
= -2 \kappa\, v(\ell)\, d{\cal A}\, .
\ee
Given that $\xi$ commutes with the Killing vector fields $\{\ell\}$, we find
similarly that 
\be
dS_{MNP_1\dots P_p}\ \xi^N D^M\left(\ell_{(1)}^{P_1} \dots
\ell_{(p)}^{P_p}\right) =0\, .
\ee
{}Finally, the Killing vector fields
${\bf m}$ are necessarily tangent to the
horizon (as for black holes) and hence orthogonal to its normals:
\be
\xi\cdot {\bf m}\big|_H =0 \, ,\qquad \ell_{(m)}\cdot {\bf m}\big|_H =
0\, .
\ee
Since ${\cal L}_\xi \ell_{(m)}=0$,
it then follows that the horizon integral involving ${\bf m}$ vanishes.
Putting together these results for the horizon integrals, and assuming that
the zeroth law of black hole mechanics (constancy of $\kappa$ on the
horizon) continues to apply to black branes, we can simplify
(\ref{newtension}) to
\be\label{presmarr}
{\cal E} = \left({n-1 \over n-2}\right)\left[ \kappa\, {\cal A}_{eff}
\ +\ \bfO_H\cdot {\bf J}\right]\, ,
\ee
where 
\be
{\cal A}_{eff} \equiv \oint_H v(\ell)\, d{\cal A}
\ee
is the `effective $(n-1)$-area' of the horizon. This generalizes the formula
of \cite{MP} for D-dimensional stationary black holes to stationary
and translationally-invariant black branes. Note that the {\it total}
horizon $(n+p-1)$-area is
\be
A = {\cal A}\,  v(\ell)V_p = {\cal A}_{eff} V_p\, .
\ee

An entirely analogous computation, applied to the Komar-type formula
(\ref{Tcov2}) for the brane tension yields the result
\be\label{tsmarr}
{\cal T} =  \left({1\over n-1}\right)\, {\cal E}\, .
\ee
An immediate consequence of this is that an uncharged
isotropic stationary black brane cannot
be boost invariant, since boost-invariance implies ${\cal T}={\cal E}$. The
qualification `black' is essential here since there is no general principle
that
would forbid uncharged boost-invariant branes; Dirac's relativistic membrane
action
and its p-brane generalization provides an effective (worldvolume)
description of such an object.

\section{Behaviour under dimensional reduction}

Given the black hole result of \cite{MP}, the Smarr-type formula
(\ref{presmarr}) for black branes is essentially what one would expect from
the fact that black holes are obtainable from black branes by dimensional
reduction, except possibly for the appearance of the factor
$v(\ell)$ in the `effective' horizon area. We will now confirm this
intuition, and find an interpretation for
${\cal A}_{eff}$ and a check on the relation (\ref{tsmarr}).

Consider the black hole solution of the $d=(n+1)$ dimensional Einstein
equations obtained by dimensional reduction of a black p-brane solution of
the $D=(d+p)$ dimensional Einstein equation along orbits of the vector
fields $\{\ell\}$. In coordinates $(x^I,y^m)$, where
$I=(0,i)$ and $\partial/\partial y^m$ are the $p$ commuting translational
Killing vector fields, the general translationally-invariant p-brane
D-metric takes the form
\be
ds^2_D = g_{IJ} dx^I dx^J + g_{mn} \left(dy^m + {}^{(p)}\!
g^{mp}g_{pK}dX^K\right) \left(dy^n + {}^{(p)}\!
g^{nq}g_{qL}dX^L\right)
\ee
where ${}^{(p)}\!g^{mn}$ is the $p\times p$ matrix inverse of $g_{mn}$,
and all components are $y$-independent. Note that
\be
v(\ell) = \sqrt{\det g_{mn}}\, ,
\ee 
and that
\be
\sqrt{-\det g_{MN}} = v(\ell)\, \sqrt{-\det g_{IJ}} \, .
\ee
We will assume that the p-brane has finite total p-volume $V_p$. Integrating
over the p-surface spanned by orbits of the Killing vector fields
$\{\ell\}$, and then dividing by $V_p$, reduces the D-dimensional
Einstein action to the $d$ dimensional one
\be
I= {1\over 2}\int d^dx\, \sqrt{-\det \tilde g}\, \tilde R \ + \tilde
I^{(mat)}\, ,
\ee
where 
\be\label{map}
\tilde g_{IJ} = \left[v(\ell)\right]^{2\over n-1} g_{IJ}\, ,
\ee
and $\tilde I^{(mat)}$ now includes the vector fields $g_{Lm}$ and the
scalar fields $g_{mn}$. It follows immediately from (\ref{map}) that
the black hole horizon $(n-1)$-area element equals $v(\ell)$ times the
$(n-1)$-area element of the p-brane horizon, which is precisely the
`effective' area density ${\cal A}_{eff}$.

It also follows from (\ref{map}) that
\be
\tilde h_{IJ} = h_{IJ} + {1\over n-1} \eta_{IJ} h_{mm} + {\cal
O}\left(h^2\right) \, .
\ee
The ADM energy on constant time surfaces of the $d$-dimensional
spacetime is
\bea
E &=& {1\over2}\oint_\infty dS_i \left[\partial_j \tilde h_{ij} - \partial_i
\tilde
h_{jj}\right]\nn
&=& {1\over2}\oint_\infty dS_i \left[\partial_j \tilde h_{ij} - \partial_i
\left(\tilde h_{jj}
+h_{mm}\right)\right]\, ,
\eea
in agreement with the ADM integral (\ref{pADM}) for $p$-brane energy
density. Similarly, the Komar energy integral for the $d$-dimensional
spacetime is equivalent to
\bea
E &=& - \left({n-1\over 2(n-2)}\right) \oint_\infty dS_i \, \partial_i
\tilde h_{00}\nn
  &=& - \left({1\over 2(n-2)} \right)\oint_\infty dS_i \, \partial_i
\left[(n-1)h_{00} -
h_{mm}\right]\, ,
\eea
in agreement with the formula (\ref{pgoo}) for $p$-brane energy density.

The above analysis is as valid for charged branes, which we will consider in
the following sections, as uncharged ones, but it is instructive to consider
the special case of uncharged black p-branes of the vacuum D-dimensional
Einstein equations. These are found by a trivial lift of stationary black
hole solutions of the vacuum $d$-dimensional Einstein equations, and their
D-metric takes the simple form
\be
ds^2_D = g_{IJ} dx^I dx^J +  dy^mdy^n \, .
\ee
It follows that $h_{mn}=0$ and hence, from (\ref{altten}) and
(\ref{pgoo}), that ${\cal E}=(n-1){\cal T}$, in agreement with
(\ref{tsmarr}). It is worth noting here that {\it charged}
black p-branes cannot be obtained from charged black holes in the quite
the same way because these are not solutions of the {\it vacuum} Einstein
equations, and even if the components of the stress tensor in the extra
dimensions are zero we still have $R_{mn}$ proportional to the trace of the
matter stress tensor. Thus, {\it charged} black p-branes will not generally
have vanishing $h_{mn}$, and hence the relation (\ref{tsmarr}) need not hold
for them. 

\section{Extension to charged branes}

We now aim to generalize the Smarr-type formula (\ref{presmarr}) to allow
for (electrically) charged branes. A charged p-brane is a source for a
$(p+1)$
form potential
$A$ with $(p+2)$-form field strength $F=dA$, so we must now allow for
a matter stress tensor of the form
\be\label{tmatstress}
T_{MN}^{(mat)} = {1\over (p+1)!}\, \left[F_{MQ_1\dots Q_{p+1}}
F_N{}^{Q_1\dots Q_{p+1}} - {1\over 2(p+2)} g_{MN}F^2\right]
\ee
where $F^2 = F_{MNP_1\dots P_p}F^{MNP_1\dots P_p}$. In this case
\bea
{\cal E}^{(mat)} &=& \left({1\over (n-2)(p+1)!}\right)\int_\Sigma
dS_{MNP_1\dots P_{p-1}} \bigg\{ \ell_{(1)}^{P_1} \dots
\ell_{(p-1)}^{P_{p-1}}\ell_{(p)}^N \nn
&& \times \left[ (n-1) F^{MR_1\dots R_{p+1}}\xi^QF_{QR_1\dots R_{p+1}}
-\left({p+1\over p+2}\right) F^2\xi^M\right]\nn
&&-\ p\  \ell_{[(1)}^{P_1} \dots \ell_{(p-1)}^{P_{p-1}}\ell_{(p)]}^Q
F_{QR_1\dots R_{p+1}}F^{MR_1\dots R_{p+1}}\xi^N \bigg\}\, .
\eea
We will assume that the field strength $F$ is invariant under
the symmetries generated by the Killing vector fields $\xi$ and
$\ell_{(m)}$, and we choose a gauge for $A$ such that
\be
{\cal L}_\xi A =0\, , \qquad {\cal L}_{(m)} A =0\, .
\ee
It then follows that
\bea
\label{lemma1}
F^{MR_1\dots R_{p+1}}\xi^Q F_{QR_1\dots R_{p+1}} &=&
-(p+1)D_P\left[ F^{MPR_1\dots R_p}\xi^QA_{QR_1\dots R_p}\right]\nn
&&+\ (p+1)\left(D_PF^{MPR_1\dots R_p}\right)\left(\xi^QA_{QR_1\dots
R_p}\right)\, ,
\eea
and similar expressions with $\xi$ replaced by a brane translation
Killing vector field $\ell$. It also
follows that
\bea\label{lemma2}
\xi^M F^2 &=& 2(p+2)D_P\left[ \xi^{[M}F^{P]Q_1\dots Q_{p+1}}A_{Q_1\dots
Q_{p+1}}\right] \nn
&&-\ (p+2) \xi^M\left(D_PF^{PQ_1\dots Q_{p+1}}\right)
A_{Q_1\dots Q_{p+1}}\, .
\eea
Using these results, and the field equation
\be\label{fieldeq}
D_MF^{MN_1\dots N_{p+1}} =0\, ,
\ee
we deduce that
\bea\label{ematint}
{\cal E}^{(mat)} &=& -{(p+2) \over 2(n-2)p!}\int_\Sigma dS_{[MNP_1\dots
P_{p-1}}D_{S]}
\bigg\{ \ell_{(1)}^{P_1}\dots \ell_{(p-1)}^{P_{p-1}}\ell_{(p)}^N\bigg[
(n-1)F^{MSR_1\dots R_p} \xi^Q \nn
\ \ \ \ \  +&&\!\!\!\!\!\!\!\!\!\! 2F^{SQR_1\dots
R_p}\xi^M\bigg]A_{QR_1\dots R_p} - p\, \ell_{[(1)}^{P_1}\dots
\ell_{(p-1)}^{P_{p-1}}\ell_{(p)]}^{Q}\xi^N F^{MSR_1\dots R_p} A_{QR_1\dots
R_p} \bigg\} \, .
\eea
This bulk integral equals the difference of
surface integrals at infinity and the horizon. If we choose
a gauge for which $A$ vanishes at infinity then the surface term at infinity
also vanishes and we are left with the following surface integral over the
horizon:
\bea\label{horizon}
{\cal E}^{(mat)} &=& {1\over 2(n-2)p!}\oint_H dS_{MNP_1\dots P_{p-1}S}
\bigg\{ \ell_{(1)}^{P_1}\dots \ell_{(p-1)}^{P_{p-1}}\ell_{(p)}^N\times\nn
&& \left[(n-1)F^{MSR_1\dots R_p} \xi^Q  + 2F^{SQR_1\dots
R_p}\xi^M\right] A_{QR_1\dots R_p} \nn
&& -\ p\ \ell_{[(1)}^{P_1}\dots
\ell_{(p-1)}^{P_{p-1}}\ell_{(p)]}^Q\xi^N F^{MSR_1\dots R_p} A_{QR_1\dots
R_p} \bigg\}\, .
\eea
We now have to evaluate this integral and interpret the result.

We begin by observing that
\be\label{horlem}
\xi_M\left( \xi^{[M} F^{NQR_1\dots R_p]}\right)\big|_H =0\, .
\ee
This can be proved as follows. Both $\xi^2$ and $R_{MN}\xi^M\xi^N$ vanish
on a Killing horizon of $\xi$, so the Einstein equation (\ref{EinR}) implies
that the (p+1)-form $i_\xi F$ has vanishing norm on the horizon. However, we
also have $i_\xi(i_\xi F)\equiv 0$. These properties of $i_\xi F$ and
$\xi$ imply that $\xi \wedge i_\xi F=0$ on the horizon
(where $\xi$ is here the 1-form constructed from the timelike
Killing vector field and the metric $g$) and hence
(since $\xi$ is null on the horizon) that
$i_\xi (\xi\wedge F)$ vanishes on the horizon, but this is just
(\ref{horlem}). 
Given the form (\ref{horsurfel}) of the horizon surface element, and
defining the set of rank $(p+3)$ antisymmetric tensors
\be
K_{(m)}^{MN_1 \dots N_{p+2}} \equiv (p+3)\, \ell_{(m)}^{[M}F^{N_1\dots
N_{p+2}]}\, ,
\ee
we can now rewrite (\ref{horizon}) as
\bea\label{horizon2}
{\cal E}^{(mat)} &=& -{1\over 2(n-2)p!}\oint_H dS_{MNP_1\dots P_{p-1}S}
\bigg\{ \ell_{(1)}^{P_1}\dots
\ell_{(p-1)}^{P_{p-1}}\ell_{(p)}^NF^{MSR_1\dots R_p} \Phi_{R_1\dots R_p}\nn
&& -\ {1\over3}(p+1) \ell_{[(1)}^{P_1}\dots \ell_{(p-1)}^{P_{p-1}}
K_{(p)]}^{NMSR_1\dots R_p} \Phi_{R_1\dots R_p} \bigg\} \, ,
\eea
where
\be
\Phi_{R_1\dots R_p} = -\xi^QA_{QR_1\dots R_p}\, .
\ee

We shall now make the simplifying assumption that
\be\label{truncate}
K_{(m)}^{M_1\dots M_{p+3}} =0 \qquad (m=1,\dots,p).
\ee
We postpone discussion of the significance of this constraint. For the
moment we need only observe that it allows us to simplify the horizon
integral (\ref{horizon2}) to
\be
{\cal E}^{(mat)} = {1\over (p+2)!} \oint_H dS_{MNP_1\dots P_p}\
F^{MNP_1\dots P_p} \Phi_H\, ,
\ee
where $\Phi_H$ is the electric potential of the horizon, given by
\be
\Phi_H = \ell_{(1)}^{R_1}\dots \ell_{(p)}^{R_p}
\Phi_{R_1\dots R_p} \big|_H \, .
\ee
The electric potential is constant over the horizon. To see this we first
recall that $i_\xi F\wedge \xi=0$ on the horizon. This implies, since
$\xi\cdot \ell_{(m)}=0$ on the horizon, that
\be
\left(i_\xi i_{(1)}\dots i_{(p)} F\right)\wedge \xi=0
\ee
on the horizon, where $i_{(m)}$ indicates contraction with the vector field
$\ell_{(m)}$. It follows that the 1-form $i_\xi i_{(1)}\dots i_{(p)} F$ is
proportional to $\xi$ on the horizon, and hence that
\be
i_t (i_\xi i_{(1)}\dots i_{(p)} F)\big|_H=0
\ee
for any tangent $t$ to the horizon. Since $A$ is assumed to be invariant
under the symmetries generated by $\xi$ and $\ell_{(m)}$ this is equivalent
to
\be
i_t d\Phi_H=0\, ,
\ee
which states that $\Phi_H$ is constant on the horizon. We can therefore take
$\Phi_H$ outside the integral to get
\be\label{presmarr2}
{\cal E}^{(mat)} = {1\over (p+2)!} \Phi_H \oint_H
dS_{MNP_1\dots P_p} F^{MNP_1\dots P_p} \, .
\ee

Given the field equation (\ref{fieldeq}) we can move the integration
surface from the horizon to infinity to get
\be
{\cal E}^{(mat)} = \Phi_H {\cal Q}
\ee
where 
\be
{\cal Q}= {1\over (p+2)!}\oint_\infty dS_{MNP_1\dots P_p} F^{MNP_1\dots
P_p}
\ee
is the total `electric' charge per unit p-volume. Using this result in
(\ref{presmarr}) we arrive at the following  Smarr-type formula for the
energy per unit p-volume of a  {\it charged} black p-brane:
\be\label{psmarr}
{\cal E} = \left({n-1 \over n-2}\right)\kappa\, {\cal A}_{eff}
\  + \ \left({n-1 \over n-2}\right) \bfO_H\cdot {\bf {\cal J}}
\ +\ \Phi_H {\cal Q} \, .
\ee
This generalizes the result of \cite{GMT} for charged black holes.
An entirely analogous calculation, applied to the Komar-type formula
(\ref{altten}) for the brane tension yields the relation
\be\label{tsmarr2}
{\cal T} - \Phi_H {\cal Q} =  \left({1\over n-1}\right)\left[ {\cal E}
- \Phi_H {\cal Q}\right] \, ,
\ee
which generalizes (\ref{tsmarr}).

In the supergravity context, the energy density of any `physical'
isotropic brane solution
of the vacuum supergravity equations can be shown to satisfy the
bound\footnote{The
proof of this bound in \cite{GHT} does not assume supersymmetry but it turns
out that, for $p>0$, the conditions required to establish it are satisfied
only
by theories that are `supersymmetrizable'. As the proof does not assume the
existence of a horizon, the coefficient of proportionality was not
identified in \cite{GHT} as the horizon potential; but this was
pointed out for D=5 black holes in \cite{GMT}.}
\be\label{bound}
{\cal E} \ge \Phi_H {\cal Q}\, ,
\ee
with equality for solutions that are `supersymmetric'. From
(\ref{psmarr}) it can be seen that this condition is equivalent to
vanishing $\kappa$ and $\bfO_H$; `supersymmetric'
is therefore a stronger condition than `extremal' ($\kappa=0$).
When combined with (\ref{tsmarr2}), the bound on the energy density
implies the bounds
\be
{\cal T} \le {\cal E}\, ,
\ee
and 
\be
{\cal T} \ge \Phi_H {\cal Q}\, ,
\ee
both of which are saturated by supersymmetric (i.e., isotropic
boost-invariant) branes.

\section{The first law of black branes}

So far we have derived a Smarr-type formula for the
energy per unit p-volume of a
stationary and translationally invariant black p-brane that
solves the Euler-Lagrange equations of the action
\be\label{action}
I = {1\over 2}\int d^Dx\, \sqrt{-\det g} \left[ R - {1\over
(p+2)!}F^2\right]\, ,
\ee
for $(p+2)$ form field strength $F=dA$. If the metric and the
$(p+1)$-form potential $A$ are assigned dimension zero
then the dimension of each term in the action is determined by the number of
derivatives. Since both terms in (\ref{action}) have two derivatives they
also have the same dimension, and it follows from this that the field
equations are invariant under a constant rescaling of the coordinates, which
therefore takes one black p-brane solution into another one via a rescaling
of any dimensionful parameters on which the solution depends. Thus any
dimensionful quantity, such as the horizon area, must be a weighted
homogeneous function of any set of independent parameters of the solution,
the weights being determined by the dimensions of these parameters.
In the black hole ($p=0$) case, the uniqueness theorems imply that the only
independent parameters of the solution are the mass $M$, charge $Q$ and
angular momenta ${\bf J}$, so the horizon area $A$ is a weighted homogeneous
function of them. Equivalently, $M$ is a weighted homogeneous function of
$Q$, ${\bf J}$ and $A$. Given this, the first law may be derived by an
application of Euler's theorem on weighted homogeneous functions\footnote{We
believe that this argument was originally due to G.W. Gibbons; dimensional
considerations also play a part in the proof of the first law developed by
Wald \cite{wald}.}. This method was used  in \cite{GMT} to derive the first
law for charged black holes in an arbitrary spacetime dimension from a
Smarr-type relation that is formally identical to (\ref{psmarr}). The only
difference is that for $p>0$ the extensive variables $M,Q,{\bf J},A$ must be
replaced by the densities ${\cal E},{\cal Q},{\bf {\cal J}}, {\cal
A}_{eff}$. The first three of these are parameters that determine the
coefficients of the asymptotic metric and $(p+1)$-form potential, and
their dimensions are therefore determined by the requirement that
the fields be dimensionless. This yields\footnote{It might seem odd
that the dimensions of ${\cal E}$ and ${\cal J}$, for example,
are not related in the way one would expect for energy and angular
momentum, but this is due to implicit factors of Newton's constant
$G$.}
\be
[{\cal Q}] = [{\cal E}] = L^{n-2}\, ,\qquad [{\cal J}] = L^{n-1}
\ee
In addition, we have
\be
[{\cal A}_{eff}] = L^{n-1}
\ee
because it is a measure of the `area' of an $(n-1)$-surface. We thus
have, for all $p$, 
\be
[{\cal Q}] = [{\cal E}] \, ,\qquad [{\cal J}] = [{\cal A}_{eff}] = [{\cal
E}]^{(n-1)/(n-2)}\, .
\ee
If we know that ${\cal E}$ is some function of ${\cal Q}$, ${\bf {\cal J}}$
and ${\cal A}_{eff}$ (which presumably follows from some extension to
$D>4$ and $p>0$ of the uniqueness theorems for D=4 black holes) then Euler's
theorem implies that
\be
{\cal E} = {\cal Q}{\partial {\cal E}\over\partial {\cal Q}} +
\left({n-1\over n-2}\right) {\bf {\cal J}}\cdot {\partial {\cal E}\over
\partial {\bf {\cal J}}} + \left({n-1\over n-2}\right) {\cal A}_{eff}
{\partial {\cal E}\over\partial {\cal A}_{eff} }\, .
\ee
Comparing this with (\ref{psmarr}) we see that
\be
{\partial {\cal E}\over\partial {\cal Q}} = \Phi_H\, ,\qquad
{\partial {\cal E}\over \partial {\bf {\cal J}}} = \bfO_H\, ,\qquad
{\partial {\cal E}\over\partial {\cal A}_{eff} } = \kappa
\ee
and hence the first law,
\be\label{firstlaw}
d{\cal E} = \kappa\,  d{\cal A}_{eff} + \Phi_H d {\cal Q} + \bfO_H\cdot
d{\bf
{\cal J}}\, .
\ee

For even $p$ and spacetime dimension $D=3p+5$ one can add to the action an
additional `$FFA$' Chern-Simons (CS) term. This leads to a modification of
the field equation (\ref{fieldeq}) to one of the form
\be
D_MF^{MN_1\dots N_{p+1}} \propto Z^{N_1\dots N_{p+1}}\, ,
\ee
where
\be
Z^{N_1\dots N_{p+1}} = \varepsilon^{N_1\dots N_{p+1}M_1\dots M_{p+2}L_1\dots
L_{p+2}}F_{M_1\dots M_{p+2}}F_{L_1\dots L_{p+2}}
\ee
is a topological $(p+1)$-form current. As the CS term has the same dimension
as the other terms in the action, the modified field equations again involve
no dimensionless parameters and the Euler's theorem argument still applies.
For $p=0$ and $p=2$ this possibility is relevant to supergravity theories
since both the pure D=5 supergravity and D=11 supergravity are of this form,
with a definite non-zero coefficient for the CS term. For $p=0$ the previous
derivation of the Smarr-type formula (\ref{psmarr}) requires modification to
take into account the fact that $D_MF^{MN}$ no longer vanishes. This was
done in \cite{GMT}, where it was shown that all the additional terms cancel.
For $p>0$ the analogous computation does not lead to a similar cancelation,
but for $p>0$ our assumption that $F$ is restricted by (\ref{truncate})
implies that the topological current $Z$ vanishes, so in either case the
formula (\ref{psmarr}) is unchanged.

At this point, we should address the physical meaning of the condition
(\ref{truncate}). It implies that
\be\label{trunc2}
F_{MNP_1\dots P_p} = f_{[MN}\ell^{(1)}_{P_1}\dots \ell^{(p)}_{P_p]}
\ee
for some 2-form $f$, so (\ref{truncate}) is essentially a truncation to
those fields that would yield a pure Einstein-Maxwell system upon
dimensional reduction to a $d=n+1$ dimensional spacetime. In view of this,
it
is not surprising that the first law of black brane mechanics
(\ref{firstlaw}) is just such that it reduces to the first law for charged
black holes of Einstein-Maxwell theory, and it could have been deduced in
this
way. However, this fact alone does not provide much insight into the status
of
(\ref{truncate}). Some insight can be had from specific examples.
Consider the rotating non-extremal membrane solutions of D=11 supergravity
of \cite{cvetic}; it may be verified that they satisfy
(\ref{truncate}). In the non-rotating case $F$ is purely electric. A
magnetic
component is induced by rotation but $F$ remains of the form
(\ref{trunc2}).  Obviously, there can be no net magnetic charge because the
magnetic object is a fivebrane. More generally, whenever $D=3p+5$ the
electric p-brane spacetime can carry no net magnetic charge, and one might
expect, by some extension of the black hole uniqueness theorems, that all
stationary and translationally-invariant p-brane spacetimes will be
determined by their energy density angular momentum density and electric
charge density. In this case, the above example suggests that
(\ref{truncate}) will be a consequence of the assumptions of stationarity
and translational invariance for actions of the form assumed here, at least
in those spacetime dimensions $D$ for which a CS term is possible.

The first law (\ref{firstlaw}) is a statement that relates a change in
the energy {\it density} to changes in other densities. By
setting
\be
E= {\cal E}V_p \, ,\qquad A= {\cal A}_{eff}V_p\, ,\qquad
{\bf J} = {\bf {\cal J}} V_p\, ,\qquad Q= {\cal Q} V_p
\ee
we deduce that
\be
dE = \kappa dA + \bfO_H\cdot d{\bf J} + \Phi_H dQ + {\cal T}_{eff} dV_p\, ,
\ee
where, after use of both (\ref{psmarr}) and (\ref{tsmarr2}), the
`effective tension' is found to be
\be
{\cal T}_{eff} = {\cal T} - \Phi_H {\cal Q}\, .
\ee
For constant $V_p$ this is just the first law of black hole mechanics,
since $A$ is equal to the total horizon area,
but a change in $V_p$ induce changes in $E$ proportional to what
we have called the `effective tension'. For {\it neutral}
branes this is just the ADM tension,
but for charged branes there is an additional, negative, contribution to
the effective tension proportional to the charge\footnote{In other words, a
positive contribution to the worldspace pressure; this is presumably due to
the
fact that a charged black brane has a non-zero stress-tensor in the region
outside
the horizon.}.

We pointed out in the previous section that ${\cal T}$ is subject to a lower
bound,
which we can now write as
\be\label{bound2}
{\cal T}_{eff} \ge 0\, .
\ee
We also pointed out that saturation of this bound occurs only for
isotropic boost-invariant branes. In all other cases, the effective tension
is
strictly
positive and the system can lower its energy by reducing $V_p$. This is
exactly what one would expect from wrapping a p-brane of positive ADM
tension on a p-torus; the surprise is that this effect of the ADM tension
is cancelled for boost-invariant branes. This can be understood from the
perspective of the effective worldvolume action. What we have called the
effective
tension can be identified as the vacuum energy of the brane. There can be a
contribution to this vacuum energy from the `Wess-Zumino' term ${\cal
Q}\int_w\!\bar A$ in the effective action, where the integral
is over the worldvolume and $\bar A$ is here the pullback to $w$ of the
background ($p+1$)-form potential.  Since the background is the Minkowski
vacuum, $\bar A$ is pure gauge.  The choice
$\bar A = -\Phi_H vol(w)$ yields a contribution of $-\Phi_H{\cal Q}$ to
the vacuum energy; this gauge choice must be the correct one because it
ensures that the vacuum energy vanishes when the worldvolume theory is
supersymmetric.

\section{Summary and Discussion}

We hope to have clarified various aspects of the ADM-type integrals for the
energy density, tension and angular momentum of `brane' spacetimes that are
not, strictly, asymptotically flat but only `transverse asymptotically
flat'; only the ADM formula for energy density had been previously
discussed. We then used these results to find new covariant Komar-type
surface integrals for the energy density, tension and angular momentum of
transverse asymptotically flat brane spacetimes that are also asymptotically
stationary and asymptotically translationally invariant in brane directions.
The formula for the energy density, in particular, is {\sl not} an obvious
generalization of the $p=0$ case because the brane tension (which
is absent for $p=0$) contributes to the energy density.

As an application of our Komar-type surface integrals, we deduced a
Smarr-type formula for the energy density of translationally invariant
charged black brane spacetimes, and thence the first law of black brane
mechanics. The simplest case of physical interest to which our analysis
applies is that of 2-brane solutions of D=11 supergravity, which couple to
the 3-form potential of D=11 supergravity. It would be of interest to extend
our results to 5-branes of D=11 supergravity but this will require a better
understanding of how to include both electric and magnetic charges in
theories with CS terms.

{}From the Smarr-type formula we deduced a version of the first law of black
brane mechanics that relates changes in densities; this is formally the same
as the first law of black hole mechanics obtained in \cite{BCH} but the
horizon area is now replaced by an `effective' area per unit p-volume. These
results can be understood via a dimensional reduction of the p-brane to a
black hole, but the first law of black hole mechanics that one obtains this
way now involves the worldspace volume and an `effective' tension as a
new conjugate pair. For uncharged black holes the effective tension equals
the ADM tension, as one might have anticipated, but for charged branes
there is an additional negative contribution that causes the effective
tension to vanish for boost-invariant branes, which are supersymmetric in
the
context of supergravity. This cancellation can be understood from the
perspective of an effective worldvolume field theory because the effective
tension
is the vacuum energy of this theory.

Although the surface integrals that we have discussed are valid for branes
that are only asymptotically translationally invariant, our derivation of
the first law assumed translation invariance. However, the first law of
black
brane mechanics must be a special case of the first law of thermodynamics,
once quantum mechanics is taken into account, so it should be possible to
derive it for branes that are not translationally invariant. Some progress
on this front has already been made in \cite{TF}; this is an important problem
because non-extremal stationary translationally invariant brane spacetimes
are generically unstable \cite{GL,GM,HR}, decaying to some stable stationary
brane that is {\it not} translationally invariant \cite{HM}. From 
the worldvolume
perspective this could be seen as a spontaneous breakdown of translational
invariance, but the worldvolume theory cannot be Poincar{\'e} invariant
because non-extremal black branes are not boost invariant.

\vskip 1cm
\noindent
{\bf Acknowledgements}: We thank Stanley Deser, Gary Gibbons, David Mateos,
Kellogg Stelle and Jenny Traschen for very helpful 
discussions and communications.

\newcommand{\NP}[1]{Nucl.\ Phys.\ {\bf #1}}
\newcommand{\AP}[1]{Ann.\ Phys.\ {\bf #1}}
\newcommand{\PL}[1]{Phys.\ Lett.\ {\bf #1}}
\newcommand{\CQG}[1]{Class. Quant. Gravity {\bf #1}}
\newcommand{\CMP}[1]{Comm.\ Math.\ Phys.\ {\bf #1}}
\newcommand{\PR}[1]{Phys.\ Rev.\ {\bf #1}}
\newcommand{\PRL}[1]{Phys.\ Rev.\ Lett.\ {\bf #1}}
\newcommand{\PRE}[1]{Phys.\ Rep.\ {\bf #1}}
\newcommand{\PTP}[1]{Prog.\ Theor.\ Phys.\ {\bf #1}}
\newcommand{\PTPS}[1]{Prog.\ Theor.\ Phys.\ Suppl.\ {\bf #1}}
\newcommand{\MPL}[1]{Mod.\ Phys.\ Lett.\ {\bf #1}}
\newcommand{\IJMP}[1]{Int.\ Jour.\ Mod.\ Phys.\ {\bf #1}}
\newcommand{\JHEP}[1]{J.\ High\ Energy\ Phys.\ {\bf #1}}
\newcommand{\JP}[1]{Jour.\ Phys.\ {\bf #1}}

\end{document}